\documentclass{article}[12pt, a4paper]

\usepackage[english]{babel}
\usepackage[dvipsnames]{xcolor}
\usepackage[a4paper,top=2cm,bottom=2cm,left=3cm,right=3cm,marginparwidth=1.75cm]{geometry}
\usepackage{authblk}

\usepackage{amsmath}
\usepackage{graphicx}
\usepackage[colorlinks=true, allcolors=blue]{hyperref}
\usepackage{cite}
\usepackage{soul}
\usepackage{array}
\usepackage{multirow}
\usepackage{subcaption}

\newcommand{\fix}[2]{{\bf FIX}\footnote{{\bf{Fix #1:} }#2}}

\newcommand{\comment}[1]{}

\comment{
   this is the way you can use them in the text
   \fix{name of author}{the note} 
}

\graphicspath{{Figures/}}
\newcommand{\PreserveBackslash}[1]{\let\temp=\\#1\let\\=\temp}
\newcolumntype{C}[1]{>{\PreserveBackslash\centering}p{#1}}

\title{Information Fusion in Multimodal IoT Systems for physical activity level monitoring}

\author[1]{Mohsen Shirali}
\author[2]{Zahra Ahmadi}
\author[3]{Jose-Luis Bayo-Monton}
\author[3]{Zoe Valero Ramon}
\author[3,4]{Carlos Fernández-Llatas}

\affil[1]{Department of Computing Science and Engineering (INGI), Université catholique de Louvain (UCLouvain), 1348 Louvain-la-Neuve, Belgium (\texttt{mohsen.shirali@uclouvain.be})}
\affil[2]{Research Centre for Information Systems Engineering (LIRIS), KU Leuven, Brussels 1000, Belgium (\texttt{zahra.ahmadi@kuleuven.be})}
\affil[3]{Process Mining 4 Health Lab–SABIEN-ITACA Institute, Universitat Politècnica de València. Valencia 46022, Spain (\texttt{jobamon@itaca.upv.es, zoevara@itaca.upv.es, cfllatas@itaca.upv.es})}

\affil[4]{Department of Clinical Sciences Intervention and Technology (CLINTEC), Karolinska Institutet, Stockholm 17177, Sweden}
\date{}

\begin{document}
\maketitle

{\small{Corresponding author: Mohsen Shirali (\texttt{m\_shirali@sbu.ac.ir}) and Zoe Valero Ramon (\textit{zoevara@itaca.upv.es})}}

\begin{abstract}
This study exploits information fusion in IoT systems and uses a clustering method to identify similarities in behaviours and key characteristics within each cluster. This approach facilitates early detection of behaviour changes and provides a more in-depth understanding of behaviour routines for continuous health monitoring.

\end{abstract}

\textbf{Keywords. IoT, Physical Activity Monitoring, Fusion.}

\section{Introduction}
\label{intro}

The study of human behaviour has gained increasing attention, particularly within health and well-being in recent years. These fields acknowledge the close connection between human behaviour patterns and health conditions in a way that an improper lifestyle or changes in daily habits can be early indicators of diseases~\cite{fleury2009svm,ma2017revealing}.
For instance, physical inactivity stands out as a significant risk factor for certain diseases such as obesity, diabetes, and cardiovascular problems, underscoring the importance of accurately measuring activity levels to identify individuals at risk~\cite{qi2020overview}.
Many research studies also have proved the utility of monitoring daily activities to detect abnormal behaviours and discover deviations from typical routines, signaling early stages of health problems (e.g. dementia, Alzheimer's, osteoporosis, arthritis)~\cite{santacruz2001early, enshaeifar2018internet,alberdi2018smart} or timely detection of hazardous incidents occurrence (such as falls)~\cite{garcia2020ambient}.

Consequently, in light of this relationship between human activities and behaviours with health conditions, global health associations and organizations regularly release guidelines to encourage people to achieve healthier physical conditions through lifestyle strengthening.
These guidelines offer specific recommendations for individuals, mainly concerning quantities and frequencies of specific activities.
In this context, by understanding people's daily activities and behaviour routines, healthcare professionals can ensure individuals engage in the right activities that are conducive to their health at the right times.
Therefore, a comprehensive insight into a patient's health condition is a prerequisite for healthcare experts, if they aim to provide effective healthcare services.

The professionals can keep track of individuals' activities to inspect whether their behaviour aligns with the health guidelines. This information also allows the implementation of more precise interventions to prevent the onset of diseases and enhance the accuracy of medical treatments~\cite{ma2017revealing}. Moreover, on a broader scale, tracking daily behaviours facilitates lifestyle monitoring, paving the way for offering personalised health solutions.

Daily behaviour refers to a recurring sequence of human movements, actions or gestures, and the patterns for these behaviours can provide insights into an individual's daily routines, habits, and activities~\cite{cook2014mining,palipana2017recent}. Examples of behaviour patterns include sleep/wake cycles, bathroom usages, meal times, and mobility routines~\cite{FRITZ2022Nurse,Shirali2020Mobility}. Hence, the most common way to study an individual's behaviour is by monitoring everyday activities. Previously, these activities were assessed manually through interviews and questionnaires, which was time-consuming and prone to errors. Luckily, the introduction of Internet of Things (IoT) technologies and the widespread use of smartphones and wearable devices have made automated monitoring feasible and accessible and benefit this process by making it more efficient and accurate~\cite{stikic2008adl,cornacchia2016survey}.

The main idea of IoT is to connect anything/everything (e.g., sensors, devices, machines, people, etc.) to gather intelligence from such objects~\cite{rayes2022internet,whitmore2015internet}.
In this way, the IoT devices enable continuous monitoring of individuals and understanding of human behaviour, by providing valuable objective data on their activities (possibly in real-time)~\cite{cornacchia2016survey, jsan1030217}.
Sensor data are collected passively without human effort; allowing users to forget about the device and continue with their day~\cite{morita2023health, sztyler2015discovery}.
Meanwhile, using data mining techniques on the collected sensor readings, different types of activities can be recognised and classified based on their unique sensor signatures and in this way behaviour patterns can be extracted.

The activities that have been explored in the literature are different from the perspective of their impact on health status, the difficulty of sensing, detecting and performing~\cite{cook2014mining, wang2018towards}.
Therefore, different systems consist of wearable and ambient devices equipped with various types of sensors are commonly used to identify activities and behavioural patterns.
Caregivers and researchers are primarily interested in monitoring the ability of individuals to perform two sets of activity classes; Activities of Daily Living (ADLs) and instrumental ADLs (iADLs), because of their role in health management and the assessment of individuals' ability for living independently~\cite{krishnan2014activity}. ADLs involve self-care tasks like bathing, eating, walking and sitting, while iADLs involve activities related to interacting with the physical and social environment, such as food preparation, housekeeping, using the telephone and managing medications~\cite{stikic2008adl}.

The main problem is that raw data from IoT devices often have inadequate quality and each sensor type comes with limitations in detecting the activities, which can hinder the full utilization of IoT systems and data mining potential~\cite{Shirali2024error}.
As a result, domain experts, such as medical staff, may not fully rely on insights generated by IoT technology and its corresponding analysis and may question the credibility of these systems~\cite{sheth2014applications}.
Moreover, the outcome of the analysis, e.g. behaviour models can be hard to comprehend due to their complexity.

In this article, we aim to investigate how using multiple sources of data and the synergy of information from multiple data sources affects the quality and comprehensiveness of behavioural modelling.
We want to determine whether this approach increases the readability and effectiveness of discovered behaviour models or adds to the complexity of the models and makes them unreadable.

The rest of the paper is organised as follows.
In section~\ref{sec:RW}, some required background knowledge and related works are mentioned.
Section~\ref{sec:Problem&Dataset} introduces the proposed approach and selected elderly care case study with detailed information on the used dataset.
Then, in~\ref{sec:Analysis} we illustrated our experiments on the use case dataset and the results for investigation of daily routines based on location and activities are represented.
Finally, Section~\ref{sec:Conclusion} concludes the paper.

\section{Related work}
\label{sec:RW}

With the advent of ambient and wearable sensors, many efforts have been made by the academia and industry to leverage IoT and human-centric technologies such as Ambient Assisted Living (AAL), smart homes and wearables in healthcare domain in order to address the increasing demand for higher quality of health and well-being services~\cite{whitmore2015internet,rayes2022internet, dohr2010internet,jsan1030217}.
These systems primarily target elderlies, patients with chronic diseases and individuals with disabilities to improve their quality of life and independence while reducing the burden on caregivers and healthcare systems. Some examples include fall detection, medication reminders, home automation, and social interaction platforms~\cite{pires2016data}.

In many studies, IoT devices with built-in sensors have been used with machine learning algorithms for continuous long-term monitoring which refers to collecting data of user’s activities on an ongoing basis, without interruption or gaps in data collection~\cite{FRITZ2022Nurse}.
Based on the collected data, the 
behaviour patterns (or markers) such as sleep/wake behaviours, activities or activity levels can be identified.
These markers were used to predict pain~\cite{fritz2020automated} or mobility, cognition, and depression symptoms in older adults~\cite{alberdi2018smart}.
Additionally, the changes in behaviour patterns may also be indicative of health events or changes in health status and provide insights into how individuals with chronic health conditions manage their health~\cite{FRITZ2022Nurse, sprint2020behavioral}. Also, the detection of changes in behaviour, resulted from treatment regimens for chronic conditions, could determine prescribed treatment regimen adherence and impact~\cite{sprint2016using}.

Process Mining (PM) is a discipline that provides a set of tools to discover human-understandable models from event logs~\footnote{Event log is a set of events within a time interval with every single event occurred at a given point in time~\cite{van2003workflow,van_der_aalst2016process}.}\cite{van_der_aalst2016process}.
In fact, the goal of process mining is to turn event data into insights and actions and its algorithms are able to maximise the understandability of the models inferred~\cite{Fernandez-Llatas2021}. 

A process is an ordered series of activities that are executed (repetitively) with the aim of achieving a specific goal. Thus, the notion of the process can be leveraged to describe most of the behaviours we adopt in our daily life~\cite{di2022you}. Therefore, it is possible to create a simplified graphical model of human activities on a daily basis by using process mining techniques on data collected from sensors. In this way, graphically understandable representations of human behaviours will be extracted and since they can easily be understood by human experts, the PM-discovered models offer valuable insights into specific behaviour patterns~\cite{ma2017revealing}.

In addition, when a reference model is available, conformance-checking techniques in PM can be applied to assess individuals’ daily routines in satisfying a particular goal, discover the deviations in doing the activities and even verify the possible ignorance of specific steps.
PM techniques support individual behaviour analysis not only for detecting behavioural changes but also to offer a human-understandable view of the real changes of a user~\cite{fernandez2013process}.

The applicability of different PM algorithms to create behaviour processes is reviewed in~\cite{ma2017revealing}.
The results of initial experiments demonstrate the usefulness of process mining techniques in creating graphical insights into human activity in a smart home environment~\cite{ma2017revealing}. However, the applicability of PM techniques on multi-modal datasets to merge data from multiple sources and to discover behaviour processes which represent the insight from all sensor modalities has not been investigated yet, and we are proposing this idea in the rest of the paper. 

\section{Behaviour analysis}
\label{sec:Problem&Dataset}

Our proposed approach aims to tackle the issue of using IoT systems for behavioural analysis.
It involves using well-known health measures to group daily data into several days with labels representing alignment with health guidelines, (like healthy and unhealthy labels).
We will then investigate the impact of different behaviour routines on pursuing a healthy regime and achieving health goals, and whether they align with health guidelines.
The behaviour routines are modelled and extracted based on the daily activities and movements of a subject, making them highly representative of the habits and characteristics of that person.

The proposed behaviour analysis approach aims to explore how understanding daily behaviour routines, using IoT-based systems, can assist healthcare providers in their assessment.

To gather information about individuals and their surroundings during daily activities which possibly affect behaviour routines, our first step involves deploying an IoT system equipped with various sensors and devices. This system, designed to be multi-modal, taps into different sources to offer a more complete understanding of an individual's actions and their environment. Picture it like having a team of sensors, each providing a unique perspective on what's happening. By aggregating these diverse viewpoints and leveraging the synergy of information, we create a holistic and comprehensive picture of the person's behaviour and surroundings. This approach allows us to capture a richer and more detailed dataset, providing a solid foundation for our behaviour analysis.

The second step involves using a specific measure that indicates the alignment with health guidelines. We sort behaviours into different categories – one (or more) categories for behaviours that comply with the guidelines, and other categories for those that may deviate. We then apply this grouping to instances of behaviour routines, like daily or weekly activities, based on the health measure values gathered in our data. By systematically comparing and studying behaviour routines over different periods, we gain insights into how these routines influence health outcomes.

Thirdly, we discover behaviour models for each set of behaviour routine instances, categorised based on the defined health measures. We consider all of the traces and their corresponding behaviour models in each category together to understand the patterns for each group better. Next, using the data gathered from our multi-modal IoT system, which captures various aspects of an individual's actions, we conduct multi-dimensional clustering. This involves looking at different types of events recorded by our system, like location, and linking them together based on their time correlation. Leveraging Process Mining techniques, particularly the PALIA algorithm, aids us in creating easy-to-understand models, in the form of TPAs.

For instance, the behaviour models can focus on the location of the person at a given time. The duration and frequency of events within these models influence the meaning of processes, shaping the corresponding behaviours and the variations. By closely examining behaviour models we gain insights into which routines contribute to the variations and which variations align more closely with physical health guidelines.

\section{Behaviour Analysis based on Physical Activity Levels}
\label{sec:Analysis}

Regular physical activity is one of the critical factors in prevention or treatment of diseases, as well as enhancing overall public health and well-being~\cite{blair2004fitness, sztyler2015discovery}.
Thus, an individual's level of physical activity can serve as a useful indicator of their lifestyle and health condition.
Therefore, to show how our proposed approach works, physical activity is considered as a parameter to determine different healthy and unhealthy groups and we aim to explore how daily behaviour routines impact reaching desirable physical activity levels for maintaining good health.
To achieve this, we investigated the physical activity levels for various days and identified the differences in daily behaviours.

\subsection{The levels for physical activity}
The number of steps taken each day 
is an appropriate indicator to measure physical activity.
We analysed the daily steps data (measured by the wristband in our dataset) to assess the level of physical activity on a daily basis.
To categorise physical activity level, we determined two thresholds based on recommended daily step counts and the range of values in our dataset. Hence, the three following levels are considered for physical activity:

\begin{itemize}
    \item \textbf{Insufficient Level.} for daily steps less than 4000,
    \item \textbf{Sufficient Level.} for daily steps between 4000 to 10000, and
    \item \textbf{Desirable Level.} for daily steps more than 10000.
\end{itemize}

Using these thresholds, we categorised daily steps into three levels and assigned a label to each day based on its level.
This allowed us to divide the days of our experiment into three groups: Insufficient (50 days), Sufficient (74 days), and Desirable (22 days), each representing a different level of physical activity.

\subsection{Discovering behavioural routines for physical activity level groups}
\label{sec:discovery_PM_models}
The PALIA is one of the process discovery algorithms that has been used successfully with Indoor Locations Systems and smart homes to analyse people's movements and for behaviour monitoring~\cite{Lull2021behaviour, fernandez2013process}.
It utilises various techniques for syntactical pattern recognition to generate a readable model of the process in the form of a formal automaton called timed parallel automaton (TPA).
In our case study, we have employed PALIA to model the processes of location events for each group.

We then clustered the days in each physical activity level group into multiple clusters based on the similarity of their process models and extracted the main behavioural characteristics of the days within each cluster.
By clustering the process models, we are able to group the most similar location patterns together, and then the models of clusters are used in association with statistical analysis to inspect the possible correlation between behaviour patterns and physical activity levels.

In this regard, the Quality Threshold Clustering (QTC) algorithm is utilised to cluster the discovered behaviour process models (the similarity and join parameters of 25\% and 5\% are used, respectively).
In the following sections, the results of discovering the location based process models and their clustering are presented.

\subsubsection{Group 1: Insufficient physical activity level clusters and main routines}
\label{sec:group1-insufficient}
This group pertains to the 50 days where the subject did not cover the required 4000 steps and her physical activity was not enough.

\begin{figure}[h]
     \centering
     \begin{subfigure}[h]{0.75\textwidth}
         \centering
         \includegraphics[width=\textwidth]{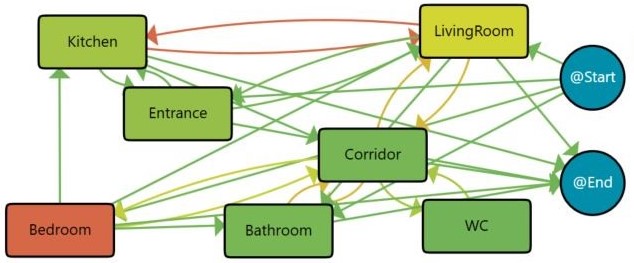}
         \caption{}
         \label{Fig:insufficient-location-basic}
     \end{subfigure}
     \begin{subfigure}[h]{0.45\textwidth}
         \centering
         \includegraphics[width=\textwidth]{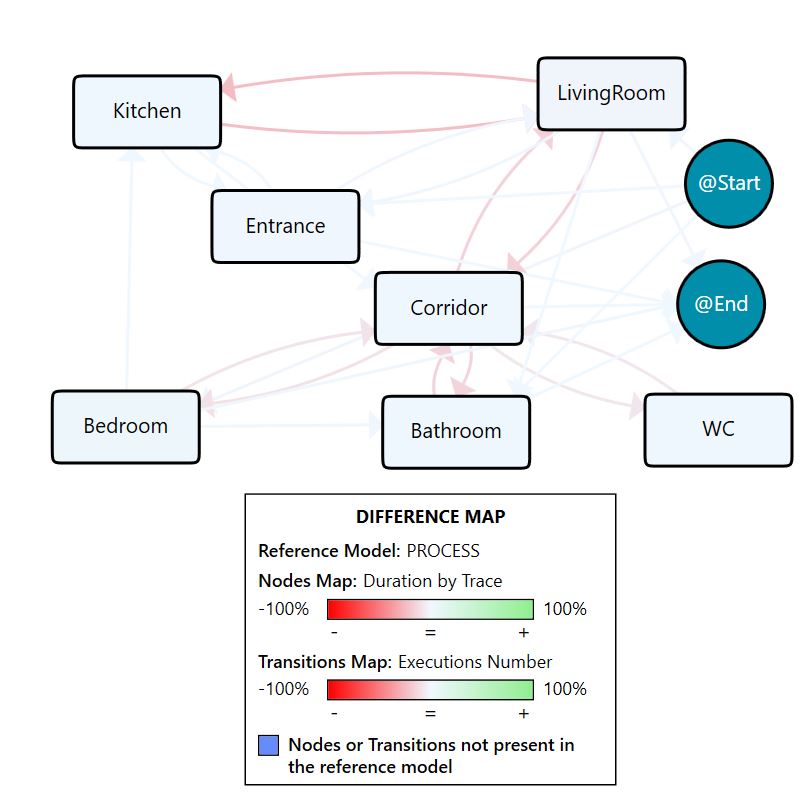}
         \caption{}
         \label{Fig:insufficient-loc1}
     \end{subfigure}
     \begin{subfigure}[h]{0.5\textwidth}
         \centering
         \includegraphics[width=\textwidth]{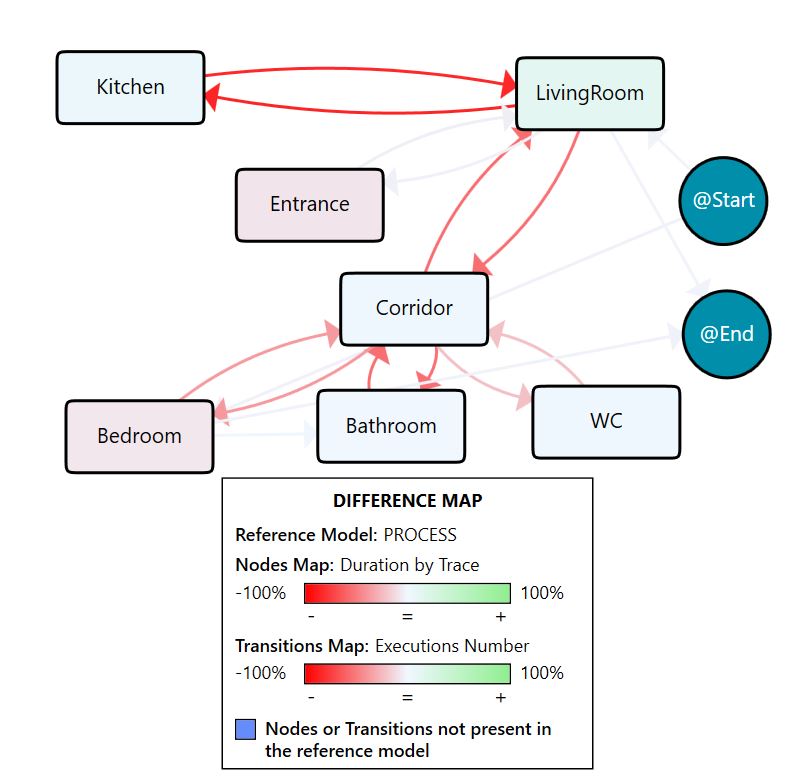}
         \caption{}
         \label{Fig:Insufficient-loc2}
     \end{subfigure}
     \caption{The TPA models for the location-based clusters of days within insufficient physical activity group; (a) The basic TPA, (b) Insufficient-L1 TPA, (c) Insufficient-L2 TPA.}
     \label{Fig:Insufficient-LocationClusters}
\end{figure}

The process model based on the location events by considering all of the 50 days within the insufficient physical activity group is illustrated in Figure~\ref{Fig:insufficient-location-basic}.
It is apparent that the person mostly spent her time in the Bedroom (half of the day) and the LivingRoom is the second most stayed place. Additionally, there were many transitions observed between the kitchen and the LivingRoom.

The location-based process maps within the insufficient physical activity level group are clustered into two main groups of 38 days and 6 days while the other 6 days are ignored and labelled as outliers due to their lack of similarity with other models. The process maps for these clusters are presented in Figure~\ref{Fig:insufficient-loc1} and~\ref{Fig:Insufficient-loc2}.
It should be noted that the nodes and transitions in the cluster maps are coloured relative to the basic location process map.

\subsubsection{Group 2: Sufficient physical activity level clusters and main routines}
The same approach for clustering based on locations is also applied to the 74 days with the label of sufficient physical activity level, the day with values between 4000 and 10000 for measured daily steps.

\begin{figure}[h]
     \centering
     \begin{subfigure}[h]{0.75\textwidth}
         \centering
         \includegraphics[width=\textwidth]{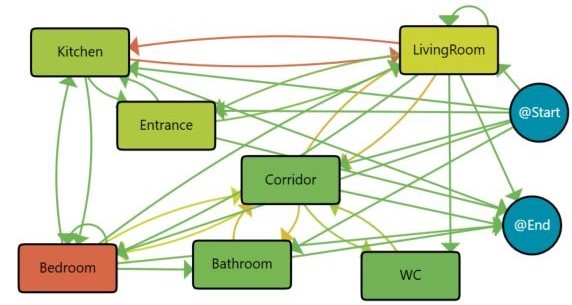}
         \caption{}
         \label{Fig:Sufficient-basic}
     \end{subfigure}
          \begin{subfigure}[h]{0.30\textwidth}
         \centering
         \includegraphics[width=\textwidth]{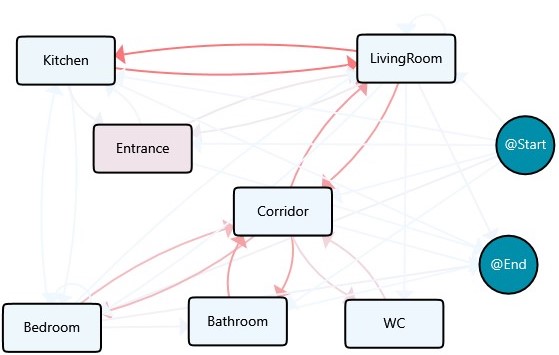}
         \caption{}
         \label{Fig:Sufficient-L1}
     \end{subfigure}
     \begin{subfigure}[h]{0.25\textwidth}
         \centering
         \includegraphics[width=\textwidth]{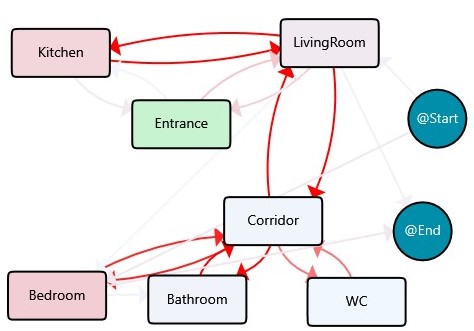}
         \caption{}
         \label{Fig:Sufficient-L2}
     \end{subfigure}
     \begin{subfigure}[h]{0.30\textwidth}
         \centering
         \includegraphics[width=\textwidth]{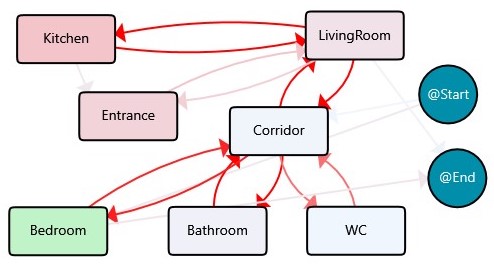}
         \caption{}
         \label{Fig:Sufficient-L3}
     \end{subfigure}
     \caption{The location-based clustered TPA models for the days within Sufficient physical activity group; (a) The basic TPA, (b) Sufficient location cluster 1, (c) Sufficient location cluster 2, (d) Sufficient location cluster 3.}
     \label{Fig:Sufficient-location-subcluster-overallFig}
\end{figure}

The extracted location-based process map for the Sufficient group is illustrated in Figure~\ref{Fig:Sufficient-basic}.
According to the TPA model, the days within the sufficient physical activity group follow the daily routines that the most time is spent in the Bedroom, LivingRoom, Entrance (outside of the home) and Kitchen respectively.
Similar to the preceding group, frequent transitions between the Kitchen and the Living Room are evident.

The location-based process maps associated with the days exhibiting sufficient physical activity levels are grouped into three primary clusters, with 53 days, 6 days, and 4 days. Conversely, the remaining 11 days are categorized as outliers. Figure~\ref{Fig:Sufficient-L1}, ~\ref{Fig:Sufficient-L2}, and \ref{Fig:Sufficient-L3} show the process maps for these clusters.

\subsubsection{Group 3: Desirable physical activity level clusters and main routines}
There are 22 days in our dataset which has more than 10000 covered daily steps and they are included in the Desirable group.

The location-based process maps associated with the physical activity level group exhibiting desirable levels are grouped into three primary clusters, with 15, 2, and 2 days and the remaining 3 days are categorized as outliers.
Figure~\ref{Fig:Sufficient-L1} (b), ~\ref{Fig:Sufficient-L2} (c), and \ref{Fig:Sufficient-L3} (d) show the process maps for these clusters.

\begin{figure}[ht]
     \centering
     \begin{subfigure}[h]{0.7\textwidth}
         \centering
         \includegraphics[width=\textwidth]{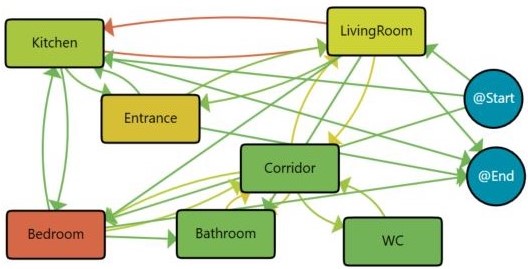}
         \caption{}
         \label{Fig:desirable-basic}
     \end{subfigure}
          \begin{subfigure}[h]{0.30\textwidth}
         \centering
         \includegraphics[width=\textwidth]{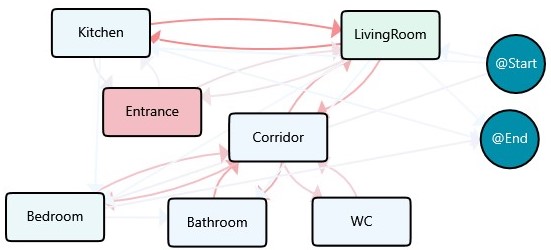}
         \caption{}
         \label{Fig:desirable-L1}
     \end{subfigure}
     \begin{subfigure}[h]{0.30\textwidth}
         \centering
         \includegraphics[width=\textwidth]{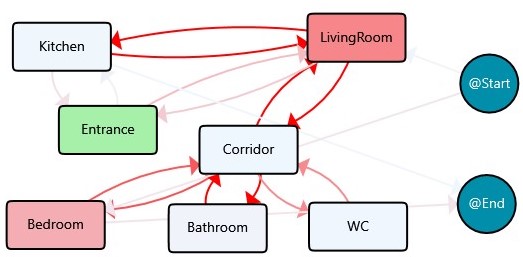}
         \caption{}
         \label{Fig:desirable-L2}
     \end{subfigure}
     \begin{subfigure}[h]{0.30\textwidth}
         \centering
         \includegraphics[width=\textwidth]{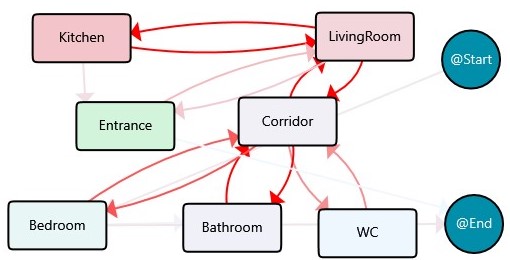}
         \caption{}
         \label{Fig:desirable-L3}
     \end{subfigure}
     \caption{The location-based clustered TPA models for the days within desirable physical activity group; (a) basic TPA, (b) TPA of Desirable-L1 cluster, (c) TPA of Desirable-L2 cluster and (d) TPA of Desirable-L3 cluster.}
     \label{Fig:desirable-location-subcluster-overallFig}
\end{figure}

Regarding the 22 days with desirable physical activity levels (i.e. the days which have more than 10000 daily steps) it is expected that the subject spend more time out of home.
As shown in Figure~\ref{Fig:desirable-basic} (a), the TPA models indicate that the days falling within this category mostly follow daily routines. These days are typically spent for approximately 10 hours in the Bedroom and around 6 hours outside the home. Similar to the previous group, there are significant movements between the Kitchen and the Living Room.

The desirable physical activity level group's associated process maps are classified into three primary clusters, which last for 15 days, 2 days, and 2 days, respectively. The residual 3 days are identified as outliers. The process maps for these clusters are displayed in Figure~\ref{Fig:desirable-L1}, ~\ref{Fig:desirable-L2}, and \ref{Fig:desirable-L3}.

\section{Conclusion}
\label{sec:Conclusion}
The paper underscores the potential of innovative behavioural analysis in understanding individual behaviours. By leveraging information from multi-modal IoT systems and aggregating the data, we can obtain a comprehensive view of physical health.
We have investigated this potential in our study by grouping daily routines in a behaviour monitoring case study.
The behaviours are modelled by process mining techniques and clustered to highlight similarities between daily routines from several days.
The outcome of behavioural analysis can provide more accurate interventions based on actual behavioural patterns.

\footnotesize
\bibliographystyle{ieeetr}
\bibliography{ref.bib}

\begin{thebibliography}{10}

\bibitem{fleury2009svm}
A.~Fleury, M.~Vacher, and N.~Noury, ``{SVM}-based multimodal classification of activities of daily living in health smart homes: sensors, algorithms, and first experimental results,'' {\em IEEE transactions on information technology in biomedicine}, vol.~14, no.~2, pp.~274--283, 2009.

\bibitem{ma2017revealing}
M.~R. Ma'arif, ``Revealing daily human activity pattern using process mining approach,'' in {\em 2017 4th International Conference on Electrical Engineering, Computer Science and Informatics (EECSI)}, pp.~1--5, IEEE, 2017.

\bibitem{qi2020overview}
J.~Qi, P.~Yang, L.~Newcombe, X.~Peng, Y.~Yang, and Z.~Zhao, ``An overview of data fusion techniques for {Internet of Things} enabled physical activity recognition and measure,'' {\em Information Fusion}, vol.~55, pp.~269--280, 2020.

\bibitem{santacruz2001early}
K.~S. Santacruz and D.~Swagerty, ``Early diagnosis of dementia,'' {\em American Family Physician}, vol.~63, no.~4, pp.~703--714, 2001.

\bibitem{enshaeifar2018internet}
S.~Enshaeifar, P.~Barnaghi, S.~Skillman, A.~Markides, T.~Elsaleh, S.~T. Acton, R.~Nilforooshan, and H.~Rostill, ``The {Internet of Things} for dementia care,'' {\em IEEE Internet Computing}, vol.~22, no.~1, pp.~8--17, 2018.

\bibitem{alberdi2018smart}
A.~Alberdi, A.~Weakley, M.~Schmitter-Edgecombe, D.~J. Cook, A.~Aztiria, A.~Basarab, and M.~Barrenechea, ``Smart home-based prediction of multidomain symptoms related to alzheimer's disease,'' {\em IEEE journal of biomedical and health informatics}, vol.~22, no.~6, pp.~1720--1731, 2018.

\bibitem{garcia2020ambient}
M.~Garcia-Constantino, A.~Konios, M.~A. Mustafa, C.~Nugent, and G.~Morrison, ``Ambient and wearable sensor fusion for abnormal behaviour detection in activities of daily living,'' in {\em 2020 IEEE International Conference on Pervasive Computing and Communications Workshops (PerCom Workshops)}, pp.~1--6, IEEE, 2020.

\bibitem{cook2014mining}
D.~J. Cook and N.~Krishnan, ``Mining the home environment,'' {\em Journal of intelligent information systems}, vol.~43, pp.~503--519, 2014.

\bibitem{palipana2017recent}
S.~Palipana, B.~Pietropaoli, and D.~Pesch, ``Recent advances in rf-based passive device-free localisation for indoor applications,'' {\em Ad Hoc Networks}, vol.~64, pp.~80--98, 2017.

\bibitem{FRITZ2022Nurse}
R.~Fritz, K.~Wuestney, G.~Dermody, and D.~J. Cook, ``Nurse-in-the-loop smart home detection of health events associated with diagnosed chronic conditions: A case-event series,'' {\em International Journal of Nursing Studies Advances}, vol.~4, p.~100081, 2022.

\bibitem{Shirali2020Mobility}
M.~Shirali, J.-L. Bayo-Monton, C.~Fernandez-Llatas, M.~Ghassemian, and V.~Traver~Salcedo, ``Design and evaluation of a solo-resident smart home testbed for mobility pattern monitoring and behavioural assessment,'' {\em Sensors}, vol.~20, no.~24, 2020.

\bibitem{stikic2008adl}
M.~Stikic, T.~Huynh, K.~Van~Laerhoven, and B.~Schiele, ``{ADL} recognition based on the combination of {RFID} and accelerometer sensing,'' in {\em 2008 second international conference on pervasive computing technologies for healthcare}, pp.~258--263, IEEE, 2008.

\bibitem{cornacchia2016survey}
M.~Cornacchia, K.~Ozcan, Y.~Zheng, and S.~Velipasalar, ``A survey on activity detection and classification using wearable sensors,'' {\em IEEE Sensors Journal}, vol.~17, no.~2, pp.~386--403, 2016.

\bibitem{rayes2022internet}
A.~Rayes and S.~Salam, {\em Internet of {Things} from {Hype} to {Reality}: {The} {Road} to {Digitization}}.
\newblock Cham: Springer International Publishing, 2022.

\bibitem{whitmore2015internet}
A.~Whitmore, A.~Agarwal, and L.~Da~Xu, ``The {Internet of Things}—a survey of topics and trends,'' {\em Information systems frontiers}, vol.~17, pp.~261--274, 2015.

\bibitem{jsan1030217}
M.~Swan, ``Sensor mania! the {Internet of Things}, wearable computing, objective metrics, and the quantified self 2.0,'' {\em Journal of Sensor and Actuator Networks}, vol.~1, no.~3, pp.~217--253, 2012.

\bibitem{morita2023health}
P.~P. Morita, K.~S. Sahu, and A.~Oetomo, ``Health monitoring using smart home technologies: Scoping review,'' {\em JMIR mHealth and uHealth}, vol.~11, p.~e37347, 2023.

\bibitem{sztyler2015discovery}
T.~Sztyler, J.~V{\"o}lker, J.~Carmona~Vargas, O.~Meier, and H.~Stuckenschmidt, ``Discovery of personal processes from labeled sensor data: An application of process mining to personalized health care,'' in {\em Proceedings of the International Workshop on Algorithms \& Theories for the Analysis of Event Data: Brussels, Belgium, June 22-23, 2015}, pp.~31--46, CEUR-ws. org, 2015.

\bibitem{wang2018towards}
A.~Wang, G.~Chen, X.~Wu, L.~Liu, N.~An, and C.-Y. Chang, ``Towards human activity recognition: A hierarchical feature selection framework,'' {\em Sensors}, vol.~18, no.~11, p.~3629, 2018.

\bibitem{krishnan2014activity}
N.~C. Krishnan and D.~J. Cook, ``Activity recognition on streaming sensor data,'' {\em Pervasive and mobile computing}, vol.~10, pp.~138--154, 2014.

\bibitem{Shirali2024error}
M.~Shirali, Z.~Ahmadi, C.~Fernández-Llatas, J.-L. Bayo-Monton, and G.~Di~Federico, ``A process mining-based error correction approach to improve data quality of an iot-sourced event log,'' {\em Under Review}, 2024.

\bibitem{sheth2014applications}
A.~Sheth, P.~Anantharam, and K.~Thirunarayan, ``Applications of multimodal physical (iot), cyber and social data for reliable and actionable insights,'' in {\em 10th IEEE International Conference on Collaborative Computing: Networking, Applications and Worksharing}, pp.~489--494, IEEE, 2014.

\bibitem{dohr2010internet}
A.~Dohr, R.~Modre-Opsrian, M.~Drobics, D.~Hayn, and G.~Schreier, ``The internet of things for ambient assisted living,'' in {\em 2010 seventh international conference on information technology: new generations}, pp.~804--809, {IEEE}, 2010.

\bibitem{pires2016data}
I.~M. Pires, N.~M. Garcia, N.~Pombo, and F.~Fl{\'o}rez-Revuelta, ``From data acquisition to data fusion: a comprehensive review and a roadmap for the identification of activities of daily living using mobile devices,'' {\em Sensors}, vol.~16, no.~2, p.~184, 2016.

\bibitem{fritz2020automated}
R.~L. Fritz, M.~Wilson, G.~Dermody, M.~Schmitter-Edgecombe, and D.~J. Cook, ``Automated smart home assessment to support pain management: multiple methods analysis,'' {\em Journal of Medical Internet Research}, vol.~22, no.~11, p.~e23943, 2020.

\bibitem{sprint2020behavioral}
G.~Sprint, D.~J. Cook, and R.~Fritz, ``Behavioral differences between subject groups identified using smart homes and change point detection,'' {\em IEEE journal of biomedical and health informatics}, vol.~25, no.~2, pp.~559--567, 2020.

\bibitem{sprint2016using}
G.~Sprint, D.~J. Cook, R.~Shelly, M.~Schmitter-Edgecombe, {\em et~al.}, ``Using smart homes to detect and analyze health events,'' {\em Computer}, vol.~49, no.~11, pp.~29--37, 2016.

\bibitem{van2003workflow}
W.~M. Van~der Aalst, B.~F. Van~Dongen, J.~Herbst, L.~Maruster, G.~Schimm, and A.~J. Weijters, ``Workflow mining: A survey of issues and approaches,'' {\em Data \& knowledge engineering}, vol.~47, no.~2, pp.~237--267, 2003.

\bibitem{van_der_aalst2016process}
W.~Van Der~Aalst, {\em Process {Mining}}.
\newblock Berlin, Heidelberg: Springer Berlin Heidelberg, 2016.

\bibitem{Fernandez-Llatas2021}
C.~Fernandez-Llatas, J.~Munoz-Gama, N.~Martin, O.~Johnson, M.~Sepulveda, and E.~Helm, {\em Process Mining in Healthcare}, pp.~41--52.
\newblock Cham: Springer International Publishing, 2021.

\bibitem{di2022you}
G.~Di~Federico and A.~Burattin, ``Do you behave always the same? a process mining approach,'' in {\em International Conference on Process Mining}, pp.~5--17, Springer, 2022.

\bibitem{fernandez2013process}
C.~Fern{\'a}ndez-Llatas, J.-M. Benedi, J.~M. Garc{\'\i}a-G{\'o}mez, and V.~Traver, ``Process mining for individualized behavior modeling using wireless tracking in nursing homes,'' {\em Sensors}, vol.~13, no.~11, pp.~15434--15451, 2013.

\bibitem{blair2004fitness}
S.~N. Blair and T.~S. Church, ``The fitness, obesity, and health equation: is physical activity the common denominator?,'' {\em Jama}, vol.~292, no.~10, pp.~1232--1234, 2004.

\bibitem{Lull2021behaviour}
J.~J. Lull, J.~L. Bayo, M.~Shirali, M.~Ghassemian, and C.~Fernandez-Llatas, {\em Interactive Process Mining in {IoT} and Human Behaviour Modelling}, pp.~217--231.
\newblock Cham: Springer International Publishing, 2021.

\end{thebibliography}

\end{document}